\documentclass{emulateapj}

\usepackage{apjfonts,graphicx,lscape,natbib}
\slugcomment{Received 2011 March 16; accepted 2011 June 23; to appear in ApJ}

\newcommand{\angstrom}{{\rm \AA}}

\newcommand{\hseventy}{{$h_{70}^{-1}$}}
\newcommand{\hbeta}{H{$\beta$}}
\newcommand{\halpha}{H{$\alpha$}}

\newcommand{\OIIIb}{[O{\sevenrm\,III}]\,$\lambda$5007}
\newcommand{\OIIIc}{[O{\sevenrm\,III}]\,$\lambda\lambda$4959,5007}

\newcommand{\NIIb}{[N{\sevenrm\,II}]\,$\lambda$6584}

 \font\sevenrm=cmr7 scaled 1000

\begin{document}


\title{Active Galactic Nucleus Pairs from the Sloan Digital Sky Survey. \\
I. The Frequency on $\sim$5--100 kpc Scales}

\shorttitle{AGN Pairs in SDSS. I}

\shortauthors{L{\footnotesize IU ET AL.}}
\author{Xin Liu\altaffilmark{1,2,4}, Yue Shen\altaffilmark{1},
Michael A. Strauss\altaffilmark{2}, and Lei
Hao\altaffilmark{3}}

\altaffiltext{1}{Harvard-Smithsonian Center for Astrophysics,
60 Garden Street, Cambridge, MA 02138}

\altaffiltext{2}{Department of Astrophysical Sciences,
Princeton University, Peyton Hall -- Ivy Lane, Princeton, NJ
08544}

\altaffiltext{3}{Shanghai Astronomical Observatory, Chinese
Academy of Sciences, 80 Nandan Road, Shanghai 200030, China}

\altaffiltext{4}{Einstein Fellow}

\begin{abstract}
Galaxy--galaxy mergers and close interactions have long been
regarded as a viable mechanism for channeling gas toward the
central supermassive black holes (SMBHs) of galaxies which are
triggered as active galactic nuclei (AGNs). AGN pairs, in which
the central SMBHs of a galaxy merger are both active, are
expected to be common from such events.  We conduct a
systematic study of 1286 AGN pairs at $\bar{z} \sim 0.1$ with
line-of-sight velocity offsets $\Delta v < 600$ km s$^{-1}$ and
projected separations $r_p < 100$ \hseventy\ kpc, selected from
the Seventh Data Release of the Sloan Digital Sky Survey
(SDSS). This AGN pair sample was drawn from 138,070 AGNs
optically identified based on diagnostic emission line ratios
and/or line widths. The fraction of AGN pairs with $5$
\hseventy\ kpc $\lesssim r_p < 100$ \hseventy\ kpc among all
spectroscopically selected AGNs at $0.02 < z < 0.16$ is 3.6\%
after correcting for SDSS spectroscopic incompleteness;
$\sim$30\% of these pairs show morphological tidal features in
their SDSS images, and the fraction becomes $\gtrsim80$\% for
pairs with the brightest nuclei. Our sample increases the
number of known AGN pairs on these scales by more than an order
of magnitude. We study their AGN and host-galaxy star formation
properties in a companion paper.
\end{abstract}

\keywords{black hole physics -- galaxies: active -- galaxies:
interactions -- galaxies: nuclei -- quasars: general}

\section{Introduction}\label{sec:intro}

Binary supermassive black holes (SMBHs) are thought to be a
generic outcome of galaxy--galaxy mergers in the hierarchical
paradigm of structure formation
\citep{begelman80,milosavljevic01,yu02}, given that most
massive galaxies harbor central SMBHs
\citep{kormendy95,richstone98}. If both SMBHs accrete material
during the same stage of a galaxy--galaxy merger, they can be
identified as a pair of active galactic nuclei (AGNs). However,
it is difficult to predict from first principles when one and
in particular both SMBHs become active during a merger
\citep[e.g.,][]{shlosman90,armitage02,wada04,dotti07}. The
frequency of AGN pairs can constrain models involving galaxy
merger rates and tidally-triggered AGN activity \citep{yu11}.
Their host-galaxy properties may offer clues to the external
and internal conditions under which both SMBHs can be activated
during galaxy--galaxy tidal encounters.

When two galaxies merge, the two SMBHs in-spiral, form a bound
binary, and harden, before they finally coalesce
\citep{begelman80}. The coalescence of binary SMBHs is expected
to be the strongest source of gravitational waves in the
universe, the detection of which would directly test general
relativity and could probe SMBH populations out to the early
universe \citep{thorne87,haehnelt94,holz05}. However, there is
still little direct observational evidence for small-separation
($\lesssim 10$ pc) gravitationally bound binary SMBHs on
Keplerian orbits (see e.g., the review of \citealt{colpi09}).
0402+379 is the only unambiguous case known. This object was
detected by Very Long Baseline Interferometry as a pair of
flat-spectrum radio point sources with a projected separation
of $\sim$7 pc (\citealt{rodriguez06,valtonen08}). At the early
stages of a merger, large-separation (kpc to tens of kpc) AGN
pairs where the galactic potential dominates can readily be
spatially resolved at cosmological distances. They provide
boundary/initial conditions for small-separation pairs. By
characterizing the large-separation population, we may gain
indirect but useful information on the small-separation
population.

There are an increasing number of examples of interacting AGN
pairs with kpc to tens-of-kpc separations. Such objects have
been identified in the X-ray
\citep[e.g.,][]{komossa03,ballo04,bianchi08,piconcelli10,mazzarella11}
the radio \citep[e.g.,][]{owen85}, or the optical
\citep[e.g.,][]{barth08,comerford09,green10,liu10b}. However,
most of the known cases of kpc-scale AGN pairs have been
serendipitous discoveries.  To systematically identify and
characterize the population of AGN pairs, we selected 167
obscured AGNs with double-peaked narrow emission lines from a
parent sample of 14,756 AGNs \citep{liu10} in the Sloan Digital
Sky Survey \citep[SDSS;][]{york00}.  Using higher resolution
near-IR imaging and spatially resolved optical spectroscopy, we
identified four kpc-scale AGN pairs among 43 of the 167
double-peaked AGNs \citep{liu10b,shen10b}.  While our work has
demonstrated that the double-peak approach allows us to
systematically identify kpc-scale AGN pairs, it has two
inherent limitations. First, the approach is biased against
pairs with line-of-sight (LOS) velocities smaller than
$\sim$150 km s$^{-1}$ due to the limited spectral resolution of
SDSS spectra. Second, it is biased against pairs with
separations $\gtrsim$ a few kpc because pairs with projected
separations larger than 3\arcsec\ (the diameter of the SDSS
fiber) will not fall within a single SDSS fiber.

To circumvent the limitations of the double-peak method and to
build up the sample of AGN pairs along the merger sequence, we
take a complementary approach to identify them.  We start with
a sample of spectroscopically identified AGNs and select
physical pairs with special attention to those that are in
mergers or close interactions.  Because each AGN in a pair has
a separate SDSS spectrum, the current approach is not limited
by the SDSS spectral resolution or by the 3\arcsec\ fiber size.
The current sample is sensitive to an earlier stage, with
separations of a few kpc to tens of kpc, of galaxy--galaxy
mergers than is the double-peak approach.  Such systems are
quite scarce, so a statistically meaningful study requires
homogeneous and large-area redshift surveys such as the SDSS.
While galaxy pairs with projected angular separations smaller
than 55\arcsec\ cannot both be observed spectroscopically on
the same SDSS plate due to the finite diameter of the fibers,
those in the overlap regions of adjacent plates can both be
spectroscopically observed \citep{strauss02,blanton03}.

The present paper is the first in the series in which we
present a statistical sample of AGN pairs from the Seventh Data
Release \citep[DR7;][]{SDSSDR7} of the SDSS. In \S
\ref{sec:selection} we describe sample selection, and discuss
incompleteness and selection biases. We address the frequency
of AGN pairs among the parent sample of optically identified
AGNs in \S \ref{sec:result}. We discuss the implications of our
results in \S \ref{sec:discuss} and conclude in \S
\ref{sec:sum}. In a companion paper (\citealt{liu11b},
hereafter Paper II), we will characterize host-galaxy recent
star formation and BH accretion properties of the AGN pair
sample presented here. Throughout we assume a $\Lambda$CDM
cosmology with $\Omega_m = 0.3$, $\Omega_{\Lambda} = 0.7$, and
$H_{0} = 70$ $h_{70}$ km s$^{-1}$ Mpc$^{-1}$.

\section{Sample Selection}\label{sec:selection}

In this section we describe our approach to select AGN pairs.
In \S \ref{subsec:parent}, we construct a parent AGN sample
optically identified from the spectroscopic sample of SDSS DR7.
We then draw AGN pairs with LOS velocity offsets $\Delta v <
600$ km s$^{-1}$ and projected separations $r_p < 100$
\hseventy\ kpc (\S \ref{subsec:pair}). To mitigate
contamination due to pairs that are closely separated but are
not interacting and to focus on AGN pairs that are
unambiguously experiencing strong tidal encounters, we visually
inspect the SDSS images of all AGN pairs and further identify a
subset, which we call the ``tidal'' sample, that exhibits
unambiguous morphological tidal features (\S
\ref{subsec:binary}). We discuss our sample completeness and
selection biases in \S \ref{subsec:completeness}.

\subsection{The Parent AGN Sample}\label{subsec:parent}

Our parent AGN sample includes objects drawn from the SDSS
``narrow-line''\footnote{The ``narrow-line'' classification
here is not meant to be strict. Some AGNs in this sample show
broad-line components as well \citep[e.g.,][]{hao05a}.} AGN
sample \citep{kauffmann03}, broad-line AGNs
\citep{hao05a,hao05b}, narrow-line quasars
\citep{zakamska03,reyes08}, and broad-line quasars
\citep{schneider10}, all from the DR7\footnote{The narrow-line
quasar sample of \citet{reyes08} was selected from a sample
corresponding to $\sim$80\% of the Data Release 6 (DR6)
spectroscopic database \citep{SDSSDR6} so we may miss a small
number of narrow-line quasars post DR6.}. Below we describe the
selection criteria for each population. The redshift cut for
all objects is $0.02 < z < 0.33$. The lower redshift limit is
to balance the need to include nearby galaxies and to avoid
redshifts where peculiar velocities can cause substantial
deviations from Hubble flow; the upper redshift limit is to
ensure \halpha\ coverage in the SDSS spectra for AGN
identification.

\begin{figure}
  \centering
    \includegraphics[width=85mm]{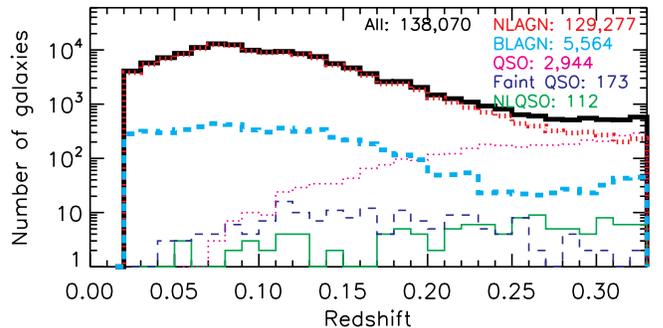}
    \caption{Redshift distributions of the parent AGN sample (black solid curve)
    and the subsamples it contains (color curves).
    We use ``NLAGN'' to denote narrow-line AGNs according to the
    \citet{kauffmann03} criterion, ``BLAGN'' for broad-line AGNs
    according to the \citet{hao05a} classification that are not
    contained in the MPA-JHU DR7 AGN catalog, ``NLQSO'' for
    narrow-line quasars in the \citet{reyes08} sample but not
    in the MPA-JHU DR7 AGN catalog, ``QSO'' for quasars in the
    catalog of \citet{schneider10}, and ``Faint QSO'' for quasars
    above the flux limit of the SDSS main galaxy sample but
    whose luminosities are too low to be included in the
    \citet{schneider10} catalog.
    Refer to \S \ref{sec:selection} for details.}
    \label{fig:z}
\end{figure}

\begin{enumerate}

\item Narrow-line AGNs. First, we select AGNs from the
    MPA-JHU SDSS DR7 galaxy
    catalog\footnote{http://www.mpa-garching.mpg.de/SDSS/DR7/}
    \citep[see][for a description of the catalog]{SDSSDR8}
    drawn from the SDSS DR7 spectroscopic database; this
    database includes 869,580 targets\footnote{Here and
    throughout, we quote the numbers of unique galaxies
    after excluding duplicate observations of the same
    galaxy.} spectrally classified as galaxies by the
    \texttt{specBS} pipeline (\citealt{SDSSDR6,SDSSDR8};
    see also discussion in \citealt{blanton05}) or quasars
    that are targeted as galaxies
    \citep{strauss02,eisenstein01} and have redshifts $z <
    0.7$. We adopt redshifts $z$ and stellar velocity
    dispersions $\sigma_{\ast}$ from the \texttt{specBS}
    pipeline. Additional spectroscopic and photometric
    properties such as emission-line fluxes and stellar
    masses are taken from the MPA-JHU data product. The
    emission line measurements are from Gaussian fits to
    continuum-subtracted spectra
    \citep{brinchmann04,tremonti04}. The MPA-JHU
    emission-line fluxes have been normalized to SDSS
    $r$-band photometric fiber magnitude and have been
    corrected for Galactic foreground extinction following
    \citet{odonnell94} using the map of \citet{schlegel98}.
    The stellar mass estimates are total stellar masses
    derived from population synthesis fits using the
    \citet{bc03} models to SDSS broadband photometry
    \citep{kauffmann03c,salim07}.

We have two selection criteria: (1) emission lines \hbeta ,
\OIIIb , \halpha , and \NIIb\ must all be detected with
signal-to-noise ratios (S/N) $> 3$, and (2) the diagnostic
emission-line ratios \OIIIb/\hbeta\ and \NIIb/\halpha\
suggest that the dominant excitation mechanism is from AGN
rather than stellar photo-ionization
\citep{bpt,osterbrock85,veilleux87} according to the
empirical criterion,
\begin{equation}
{\rm log([O\,\,{\tiny III}]}/{\rm H}\beta) > \frac{0.61}{{\rm
log([N\,\,{\tiny II}]/H}\alpha)-0.05}+1.3,
\end{equation}
as suggested by \citet{kauffmann03}. We address the effect
of the adopted S/N threshold for emission-line detections
on our results in \S \ref{sec:result}. 129,277 objects meet
these two criteria.  The \citet{kauffmann03} empirical
criterion includes AGN--H {\tiny II} ``composite'' objects,
which represent over half the sample.  If we were to adopt
instead the theoretical ``starburst limit'' suggested by
\citet{kewley01},
\begin{equation}{\rm log([O\,\,{\tiny III}]}/{\rm H}\beta) > \frac{0.61}{{\rm
log([N\,\,{\tiny II}]/H}\alpha)-0.47}+1.19,
\end{equation}
in order to exclude the AGN--H {\tiny II} composites, the
sample would be reduced to 50,624 objects.  We include
composites in our analysis, because the results would be
otherwise biased against starburst-dominated systems with
moderate AGN activity. In a study of hard X-ray selected
moderate-luminosity AGNs at $z < 0.05$ from the {\it Swift}
Burst Alert Telescope AGN sample \citep{tueller10},
\citet{koss10} find that 33\% of AGNs in merging systems
show composite or star-forming diagnostic lines. Similarly
\citet{goulding09} found evidence of star formation in the
optical emission lines of AGNs identified based on {\it
Spitzer} infrared spectra of IR luminous galaxies.

\item Narrow-line quasars. Second, we supplement the parent
    sample with 112 objects from the narrow-line quasar
    sample of \citet{reyes08}. These objects satisfy our
    selection criteria for narrow-line AGNs but are not
    already contained in the MPA-JHU SDSS DR7 galaxy
    catalog because they were not part of the main
    spectroscopic galaxy sample.

\item Broad-line AGNs. Third, we supplement the sample with
    broad-line AGNs selected by \citet{hao05a} extended to
    SDSS DR7.  These are objects selected from the SDSS
    main galaxy sample \citep{strauss02} at $z<0.33$ which
    have a broad \halpha\ component (FWHM $> 1200$ km
    s$^{-1}$) from multi-Gaussian fits with a rest-frame
    \halpha\ equivalent width (EW) $> 3$ \angstrom . There
    are 9573 such objects at $0.02<z<0.33$, and 3710 of
    them are included in the MPA-JHU DR7 AGN catalog or in
    the SDSS DR7 quasar catalog (see below). \citet{hao05a}
    show that the narrow-line components of most broad-line
    AGNs satisfy the \citet{kewley01} criteria for
    narrow-line AGNs.  Given the purpose of this work, we
    supplement the parent AGNs with the new 5863 broad-line
    AGNs, and keep the 3710 overlaps in our ``narrow-line''
    AGN category without labeling them more carefully. As
    shown by \citet{reyes08}, it can be unclear whether an
    AGN is a narrow-line or a broad-line object when the
    broad component is relatively faint or when the
    forbidden emission lines have non-Gaussian profiles
    (such as extended wings and/or double peaks). The
    classification can be further complicated by the
    presence of scattered broad-line region light in
    obscured quasars \citep[e.g.,][]{zakamska06,liu09}.

\item Broad-line quasars. We next supplement the parent
    AGNs with 2944 objects at $0.02 < z < 0.33$ from the
    SDSS DR7 quasar catalog \citep{schneider10}. The SDSS
    quasar catalog of \citet{schneider10} has a luminosity
    cut $M_i < -22.0$. There are quasars whose luminosities
    are too low to be included in the \citet{schneider10}
    quasar catalog but have extinction-corrected Petrosian
    $14.5 < r < 17.77$ mag consistent with the flux limit
    of the SDSS main galaxy sample \citep{strauss02}. To
    include those as well we searched through all objects
    targeted as ``quasars'' \citep{richards02} at $0.02 < z
    < 0.33$ in the spectroscopic sample of SDSS DR7 and
    further supplemented the sample with 173 broad-line
    objects after rejecting stellar contaminants. These
    quasars are not included in the DR7 version of the
    broad-line AGN sample as they were not targeted as
    ``galaxies'' \citep{richards02}.

\end{enumerate}

\begin{figure}
  \centering
    \includegraphics[width=85mm]{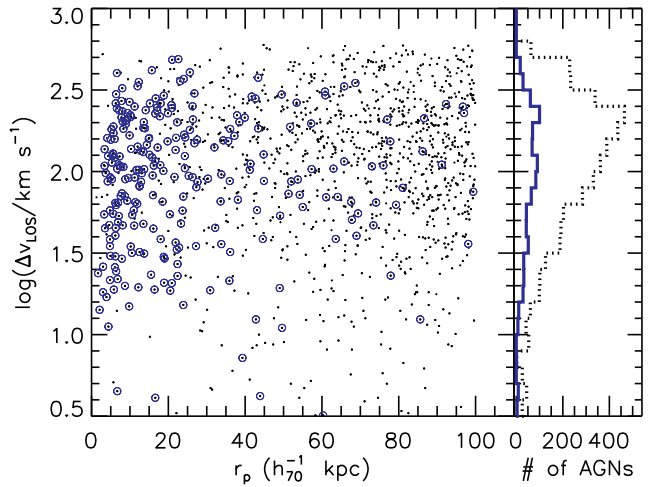}
    \caption{Projected separation $r_p$ vs. LOS velocity offset $\Delta v$
    (on a logarithmic scale) for AGN pairs (black dots and dotted histogram) and
    the subset with identified morphological tidal features (blue open circles
    and solid histogram). The $\Delta v$ distributions have been corrected for
    spectroscopic incompleteness (\S \ref{subsubsec:fiber}). We caution that
    values below $\Delta v = 30$ km s$^{-1}$ are dominated by redshift measurement
    uncertainties.}
    \label{fig:dv}
\end{figure}

The resulting parent AGN sample consists of 138,070 optically
selected AGNs at $0.02 < z < 0.33$. We list the numbers of
galaxies in each selection category in Table \ref{table:sum}.
Among the parent AGN sample, 94\% are included in the MPA-JHU
DR7 galaxy catalog, 2.9\% of which are classified as broad-line
AGNs according to \citet{hao05a} because they have weak FWHM
$>1200$ km s$^{-1}$ \halpha\ components. This catalog serves as
the starting point from which we draw the ``pair'' sample based
on projected and radial separation criteria, and a subset which
shows tidal features in SDSS images. To put our results on AGN
pairs into context and to compare with ``ordinary'' AGNs, we
also draw control AGNs from this parent sample (Paper II).
Figure \ref{fig:z} shows the redshift distribution of the
parent AGN sample and its sub-populations.

\begin{figure}
  \centering
    \includegraphics[width=85mm]{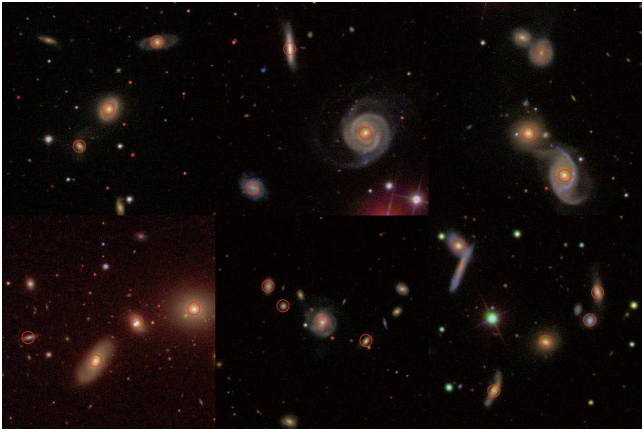}
    \caption{SDSS $gri$-color composite images \citep{lupton04}
    of six examples of AGN multiples.
    The top row displays three systems of AGN triples,
    whereas the bottom row presents two quadruples and a quintuple.
    Red circles indicate the locations of AGNs.
    North is up and east is to the left. The FOV of
    each panel is 100\arcsec\ $\times$ 100\arcsec .
    See Table \ref{table:sample} for coordinates, redshift,
    and other photometric and spectroscopic measurements of each AGN.
    }
    \label{fig:multiple}
\end{figure}

\subsection{AGN Pairs}\label{subsec:pair}

\begin{figure}
  \centering
    \includegraphics[width=85mm]{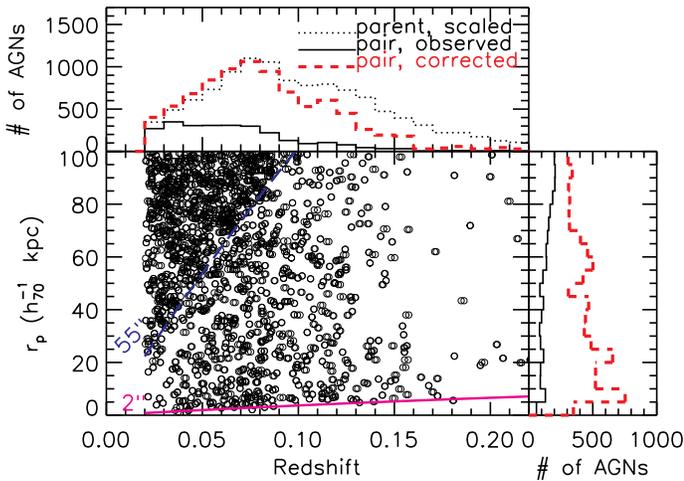}
    \caption{Redshift vs. projected separation ($r_p$) for objects in AGN pairs.
    The blue dashed curve marks the separation of 55\arcsec , below
    which the observed pair sample is incomplete due to SDSS
    fiber collision (\S \ref{subsubsec:fiber}).
    The black solid histograms show the observed redshift and $r_p$ distributions of AGN
    pairs, whereas the red dashed histograms represent the distribution
    after correction for SDSS spectroscopic incompleteness.
    The dotted histogram is the redshift distribution of the parent AGN sample scaled
    down by an arbitrary factor for display purposes.
    The magenta solid curve denotes the separation of 2\arcsec ,
    below which the observed pair sample is incomplete because
    smaller separation pairs are not resolved by SDSS imaging \citep{SDSSDR7}.}
    \label{fig:rppair}
\end{figure}

We select AGN pairs from our parent sample of AGNs based on the
following criteria: (1) projected separation $r_p < 100$
\hseventy\ kpc and (2) LOS velocity difference $\Delta v < 600$
km s$^{-1}$. We choose 100 kpc to minimize contamination due to
pairs that are closely separated but are not interacting while
maintaining a large dynamical range in separation. 100 kpc is
somewhat larger than the typical threshold values adopted in
galaxy pair studies
\citep[e.g.,][]{barton00,ellison08,darg10a}.  The
velocity-offset threshold also balances contamination and
statistics; it is comparable to the galaxy pairwise velocity
dispersion for projected separations $0.15 \leq r_p \leq 5$
\hseventy\ Mpc \citep[e.g.,][]{zehavi02}. While galaxy pairs
with small projected separations and small velocity offsets are
likely to be bound and merge within a Hubble time, the maximum
velocity offset of an interacting galaxy pair depends on local
environment. To check whether we missed a significant
population of interacting galaxies with even larger velocity
offsets, we show the distribution of $\Delta v$ of the AGN
pairs in Figure \ref{fig:dv}.  The distribution function drops
sharply at $\Delta v \gtrsim 300$ km s$^{-1}$ and the threshold
value 600 km s$^{-1}$ is on the tail of the distribution,
verifying that the $\Delta v$ threshold is adequate.

We visually inspected the SDSS images of all spectroscopically
selected pairs and rejected 34 false positives due to multiple
spectroscopic observations of different locations of the same
galaxy. The final AGN pair sample includes 1286 systems with
1244 doubles, 39 triples, two quadruples, and one quintuple
totalling 2616 AGNs.  Figure \ref{fig:multiple} shows the SDSS
images of six examples of AGN multiples, although they all have
large separations without obvious tidal features.  Using
spatially resolved spectroscopic follow-up of AGN pairs in the
sample with $r_p < 10$ \hseventy\ kpc and having a third close
companion, \citet{liu11c} discover a kpc-scale triple AGN. In
Table \ref{table:sample}, we list basic photometric and
spectroscopic measurements for all 1286 systems of AGN pairs or
multiples. Figure \ref{fig:rppair} shows their distributions of
redshift and projected separations.  While the parent sample
spans a redshift range of $0.02 < z < 0.33$, only 24 pairs are
found at $z> 0.16$ mainly reflecting the sparseness of the
parent AGN sample. To mitigate uncertainties due to small
number statistics at the higher redshift end, we restrict our
statistical analysis to the 2568 AGNs at $z < 0.16$.

\begin{figure}
  \centering
    \includegraphics[width=85mm]{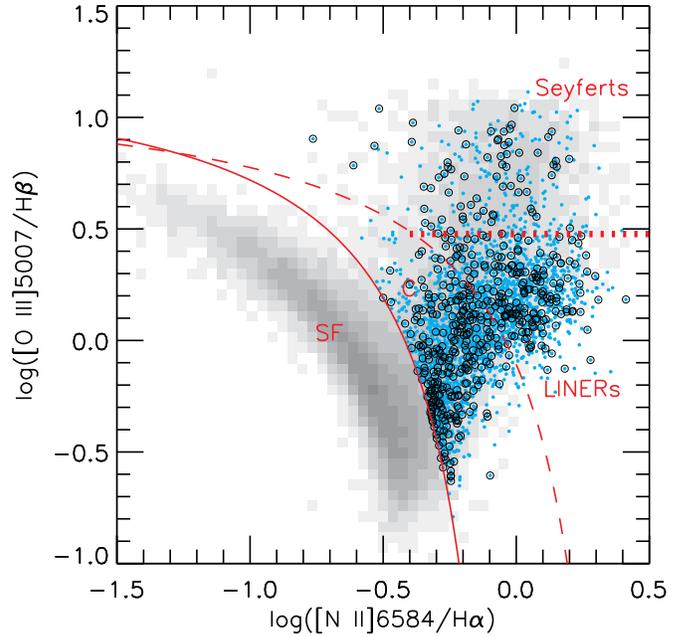}
    \caption{Diagnostic line ratios for AGNs in the pair sample (cyan dots) and
    those with tidal features (open circle).  Gray scales indicate number
     densities of 31,179 emission-line galaxies from the SDSS DR4
     \citep{kauffmann03}. The solid curve displays the empirical separation between
     H {\tiny II} regions and AGNs \citep{kauffmann03}, the
     dashed curve denotes the theoretical
     ``starburst limit'' from \citet{kewley01}, and the dotted line
     represents the empirical division between Seyferts and LINERs \citep{ho97}.
     Pure star-forming (``SF'') galaxies lie below the solid curve,
     AGN-dominated objects (Seyferts above and LINERs below the dotted line)
     lie above the dashed curve,
     and AGN-H {\tiny II} composites (``C'') lie in between.}
    \label{fig:bpt}
\end{figure}

We show in Figure \ref{fig:bpt} the optical diagnostic line
ratios \OIIIb /\hbeta\ and \NIIb /\halpha\ of AGNs in the pair
sample that are contained in the MPA-JHU SDSS DR7 galaxy
catalog.  Sixty-three percent AGNs in the pair sample lie above
the theoretical ``starburst limit'' of \citet{kewley01}. In
Paper II, we will use this sample to examine star formation and
accretion properties in the host galaxies of AGN pairs and
their correlations with pair separation.

\begin{deluxetable*}{cccccrrrccc}
\tabletypesize{\footnotesize}
\tablecolumns{13}
\tablewidth{0pc} \tablecaption{SDSS AGN
Pairs\label{table:sample} } \tablehead{ \colhead{} & \colhead{}
& \colhead{} & \colhead{} & \colhead{} & \colhead{$\Delta
\theta$} & \colhead{$r_p$} & \colhead{$\Delta v$} &
\colhead{$r$} & \colhead{} &
\colhead{}  \\
\colhead{~~~~~~~~~~~~SDSS Designation~~~~~~~~~~~~} &
\colhead{Plate} & \colhead{Fiber} & \colhead{MJD} &
\colhead{Redshift} & \colhead{($''$)} & \colhead{(\hseventy\
kpc)} & \colhead{(km s$^{-1}$)} &
\colhead{(mag)} & \colhead{F$_{{\rm AGN}}$} & \colhead{F$_{{\rm tidal}}$} \\
\colhead{(1)} & \colhead{(2)} & \colhead{(3)} & \colhead{(4)} &
\colhead{(5)} & \colhead{(6)} & \colhead{(7)} & \colhead{(8)} &
\colhead{(9)} & \colhead{(10)} & \colhead{(11)}
}
\startdata
 J00:02:49.07$+$00:45:04.8\dotfill & 0388 & 345 & 51793 & $0.0868$ &  5.9 &  9.5 &  63 &
 16.08 & 0 & 3 \\
 J00:02:49.44$+$00:45:06.7\dotfill & 0685 & 593 & 52203 & $0.0865$ &  5.9 &  9.5 &  63 &
 16.43 & 1 & 3 \\
 J00:02:57.21$+$00:07:50.5\dotfill & 0685 & 531 & 52203 & $0.0901$ & 45.4 & 76.0 & 109 &
 16.15 & 2 & 2 \\
 J00:02:58.59$+$00:08:31.0\dotfill & 0387 & 072 & 51791 & $0.0897$ & 45.4 & 76.0 & 109 &
 16.67 & 2 & 2 \\
 J00:03:23.74$+$01:05:47.3\dotfill & 1490 & 321 & 52994 & $0.0993$ & 12.2 & 22.4 &  22 &
 17.98 & 2 & 2 \\
 J00:03:23.74$+$01:05:59.5\dotfill & 0387 & 620 & 51791 & $0.0994$ & 12.2 & 22.4 &  22 &
 17.42 & 2 & 2 \\
 J00:04:25.78$-$09:58:54.4\dotfill & 0650 & 598 & 52143 & $0.1122$ & 24.1 & 49.1 &  53 &
 16.41 & 0 & 2 \\
 J00:04:26.66$-$09:58:34.4\dotfill & 0651 & 392 & 52141 & $0.1120$ & 24.1 & 49.1 &  53 &
 16.57 & 2 & 2 \\
 J00:04:31.92$-$01:14:11.7\dotfill & 0388 & 282 & 51793 & $0.0887$ & 42.3 & 70.1 &  45 &
 17.22 & 0 & 0 \\
 J00:04:33.25$-$01:13:34.4\dotfill & 1490 & 241 & 52994 & $0.0889$ & 42.3 & 70.1 &  45 &
 18.47 & 2 & 0 \\
\enddata
\tablecomments{The full table is available in the electronic
version of the paper.  Col.(1): SDSS names with J2000
coordinates given in the form of ``hhmmss.ss+ddmmss.s'';
Col.(2): Spectroscopic plate number; Col.(3): Fiber ID;
Col.(4): modified Julian date; Col.(6): angular separation;
Col.(7): transverse proper separation; Col.(8): line-of-sight
velocity offset; Col.(9): SDSS $r$-band model magnitude
corrected for Galactic extinction; Col.(10): flag for
sub-populations in the parent AGN sample (refer to \S
\ref{sec:selection} for details). ``0'' through ``2'' are for
narrow-line AGNs contained in the MPA-JHU DR7 catalog, where
``0'' stands for Seyferts, ``1'' for LINERs, and ``2'' for
composites, respectively, according to the \citet{kewley01} and
\citet{kauffmann03} criteria for separating AGNs and composites
from H {\tiny II} regions, and the \citet{ho97} criterion for
separating Seyferts from LINERs (Figure \ref{fig:bpt}). ``3''
for narrow-line quasars in the \citet{reyes08} sample but not
in the MPA-JHU DR7 catalog, ``4'' for broad-line quasars, and
``5'' for broad-line AGNs according to the \citet{hao05a}
selection; Col. (11): flag from visual identification of tidal
features (\S \ref{subsec:binary}). ``0'' for
``non-interaction'', ``1'' for ``ambiguous'', ``2'' for
``tidal'', and ``3'' for ``dumbbell'' systems.}
\end{deluxetable*}

\begin{figure*}
  \centering
    \includegraphics[width=130mm]{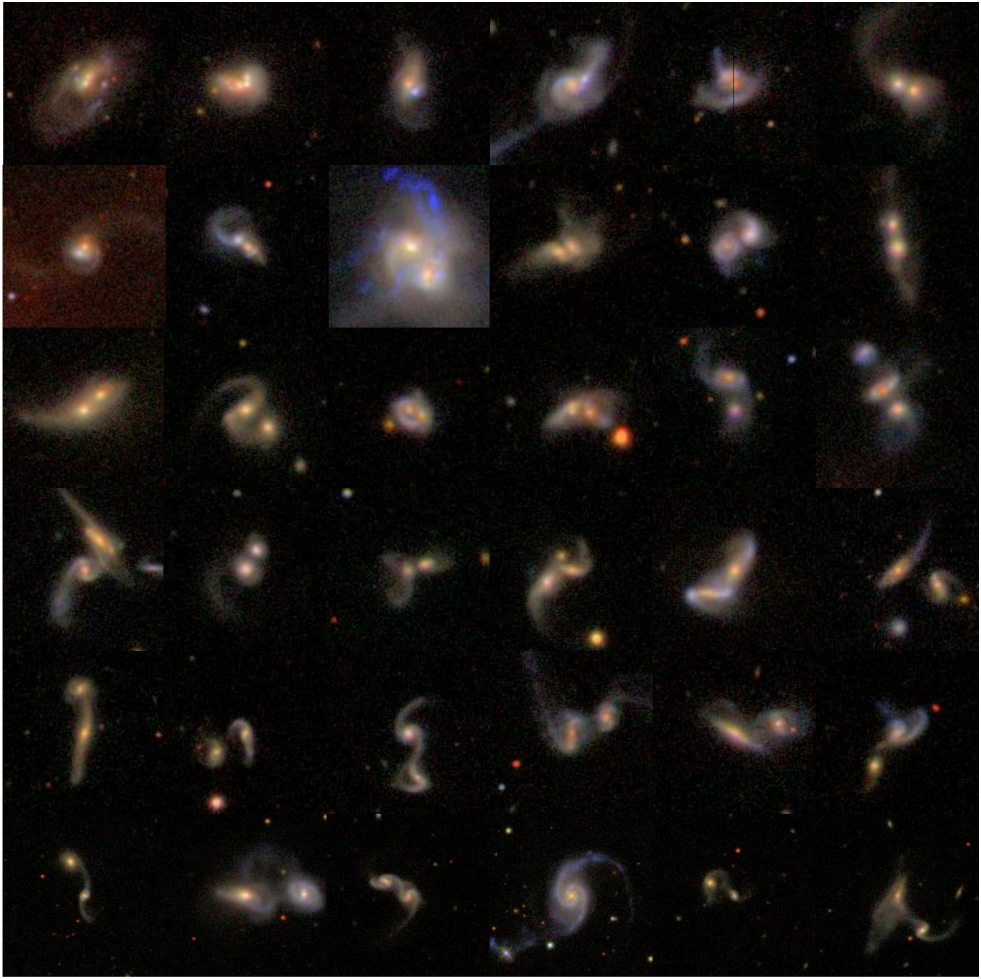}
    \caption{SDSS $gri$-color composite images of 36 examples of AGN pairs with tidal
    features.
    North is up and east is to the left.
    The FOV of each panel is 50\arcsec\ $\times$ 50\arcsec , except for those in the last
    row
    whose FOV is 100\arcsec\ $\times$ 100\arcsec .  Objects are arranged in increasing
    order of transverse proper separation $r_p$ from left to right and from top to bottom.
    Binary AGNs in the first, second, third, fourth, fifth, and last row have
    $r_p < 5$, $5 <r_p < 10$, $10 <r_p < 15$, $15 <r_p < 20$, $20 <r_p < 30$,
    and $30 < r_p < 100$ \hseventy\ kpc, respectively.
    The median redshift of the sample is 0.08.}
    \label{fig:best}
\end{figure*}

\subsubsection{Correction for SDSS Spectroscopic
Incompleteness}\label{subsubsec:fiber}

Our sample of AGN pairs is incomplete for pairs with projected
angular separations smaller than 55\arcsec\ because both
galaxies are required to have SDSS spectroscopic observations.
Due to the finite size of the SDSS fibers, galaxy pairs with
separations smaller than 55\arcsec\ will not both be observed
unless they fall in the overlap regions on adjacent plates
\citep{strauss02,blanton03}.  We correct for the fiber
collision incompleteness by supplementing the observed AGN
pairs with $(C-1)N_{p}$ systems randomly drawn from those with
separation smaller than 55\arcsec , where $N_{p}$ is the
observed number of pairs. Neglecting the tiny fraction of
triples and higher multiples, the ratio between the intrinsic
and the observed pair fraction is
\begin{equation}\label{eq:fiber}
C(f_{{\rm lap}}, p) = \frac{(1+2p)f_{{\rm lap}}f_{{\rm
spec,lap}}+(1+p)(1-f_{{\rm
lap}})f_{{\rm spec}}}{(1+2p)f_{{\rm
lap}}f^2_{{\rm spec,lap}} },
\end{equation}
where $f_{{\rm lap}} \sim$ 30\% is the plate overlap fraction
on the sky \citep{SDSSDR0}, $p$ is the ratio of pairs to
isolated galaxies, $f_{{\rm spec,lap}} \approx 0.99$ is the
SDSS spectroscopic completeness in the plate overlap region,
and $f_{{\rm spec}} \approx 0.92$ is the spectroscopic
completeness in the un-overlapped region \citep{blanton03}.
$C(f_{{\rm lap}}, p) \approx 1/f_{{\rm lap}} = 3.3$ when $p \ll
1$. We show in Figure \ref{fig:rppair} the distributions of
redshift and projected separations of AGN pairs after
correction for SDSS spectroscopic incompleteness. The corrected
redshift distribution of the pair sample is similar to that of
the parent AGNs, except that the pair sample is incomplete for
projected angular separation $\lesssim 1.''4$ due to the
imaging resolution limit of SDSS (the median PSF FWHM of SDSS
DR7 images is $1.''4$ in $r$; \citealt{SDSSDR7}). The corrected
$r_p$ distribution function increases with decreasing $r_p$
(Figure \ref{fig:rppair}).

\subsection{AGN Pairs with Tidal Features}\label{subsec:binary}

\begin{deluxetable*}{ccrrrrr}
\tabletypesize{\footnotesize}
\tablecolumns{13}
\tablewidth{0pc}
\tablecaption{Sample Summary\label{table:sum}}
\tablehead{
\colhead{~~~~~~~~~~~~~~~~~~Selection~~~~~~~~~~~~~~~~~~} &
\colhead{Sub-sample} &
\colhead{Parent} &
\colhead{Pair} &
\colhead{Tidal} &
\colhead{Ambiguous} &
\colhead{Dumbbell} \\
\colhead{(1)} &
\colhead{(2)} &
\colhead{(3)} &
\colhead{(4)} &
\colhead{(5)} &
\colhead{(6)} &
\colhead{(7)}
}
\startdata
$0.02<z<0.33$\dotfill& NLAGN        & 129277 & 2529 & 501 & 327 & 20   \\
             & NLAGN-Kewley &  50624 & 1583 & 312 & 190 & 18   \\
             & BLAGN        & ~~5564 & ~~79 & ~~9 & ~10 & ~0   \\
             & NLQSO        & ~~~112 & ~~~1 & ~~1 & ~~0 & ~0   \\
             & QSO          & ~~3117 & ~~~7 & ~~1 & ~~2 & ~0   \\
\hline
$0.02<z<0.16$, observed\dotfill & NLAGN        & 112242 & 2488 & 482 & 322 & 20   \\
                 & NLAGN-Kewley &  44168 & 1555 & 299 & 186 & 18   \\
                 & BLAGN        & ~~4595 & ~~78 & ~~8 & ~10 & ~0   \\
                 & NLQSO        & ~~~~25 & ~~~0 & ~~0 & ~~0 & ~0   \\
                 & QSO          & ~~~279 & ~~~2 & ~~0 & ~~1 & ~0   \\
\hline
$0.02<z<0.16$, corrected\dotfill & NLAGN        & 112242 & 8116 & 2487 & 1377 & 118   \\
                 & NLAGN-Kewley &  44168 & 5035 & 1552 & 802 & 106   \\
                 & BLAGN        & ~~4595 & ~236 & ~42 & ~33 & ~0   \\
                 & NLQSO        & ~~~~25 & ~~~0 & ~~0 & ~~0 & ~0   \\
                 & QSO          & ~~~279 & ~~11 & ~~0 & ~~5 & ~0   \\
\enddata
\tablecomments{Col.(1): selection criteria. The top block is
the basic selection, the middle block is restricted to $z<0.16$
to mitigate uncertainties due to small number statistics at
higher redshifts, and the bottom block is after correction for
incompleteness due to fiber collisions (\S
\ref{subsubsec:fiber}); Col.(2): subsamples in our selection of
the parent AGN sample. ``NLAGN'': narrow-line AGNs according to
the \citet{kauffmann03} criterion which is our default,
``NLAGN-Kewley'': narrow-line AGNs according to the
\citet{kewley01} criterion, ``BLAGN'': broad-line AGNs,
``NLQSO'': narrow-line quasars in the \citet{reyes08} sample
but not in the MPA-JHU DR7 catalog, ``QSO'': broad-line
quasars. Refer to \S \ref{sec:selection} for details of each
population; Col.(3): number of AGNs in the parent AGN sample;
Col.(4): number of AGNs in the pair sample (\S
\ref{subsec:pair}); Cols.(5)--(7): number of AGNs in the tidal,
ambiguous, and dumbbell categories according to our visual
classification of the SDSS images. For details of each
category, refer to \S \ref{subsec:binary}.}
\end{deluxetable*}

To mitigate contamination due to pairs that are closely
separated but are not interacting and to focus on AGN pairs
that are unambiguously experiencing strong tidal encounters, we
further select from the AGN pair sample what we dub the
``tidal'' AGN pair sample by visually examining their SDSS
images for optical tidal features such as bridges, tails,
shells, or rings. We examine both the $gri$-color composite
images \citep{lupton04} and the calibrated FITS images in all
five SDSS bands.  We use visual identification as it should be
more sensitive to low-surface-brightness (LSB) features and
less subject to false positives than automated techniques like
model fitting.  While quantitative measures of mergers are more
objective and yield reproducible results
\citep[e.g.,][]{conselice03,lotz04}, they may introduce biases
toward certain merger stages \citep[e.g., the first pericenter
passage and the final coalescence;][]{lotz08b}.

\begin{figure}
  \centering
    \includegraphics[width=85mm]{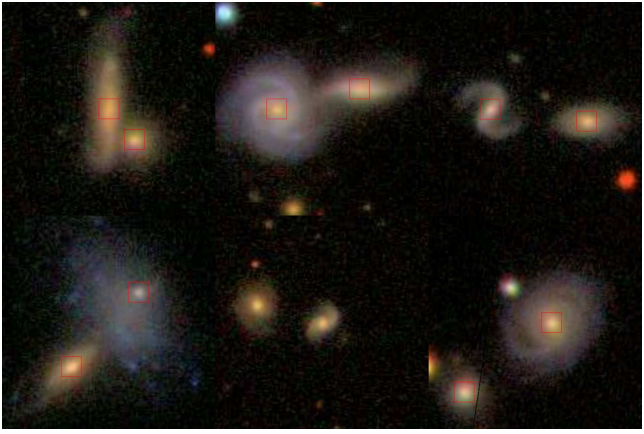}
    \caption{Examples of ambiguous systems in the pair sample. Objects
    are arranged in increasing order of transverse proper separation
    $r_p$ from left to right and from top to bottom. The FOV of
    each image tile is 50\arcsec\ $\times$ 50\arcsec . We mark the
    targets with red boxes when additional objects are within the
    FOV shown.}
    \label{fig:ambiguous}
\end{figure}

\begin{figure}
  \centering
    \includegraphics[width=85mm]{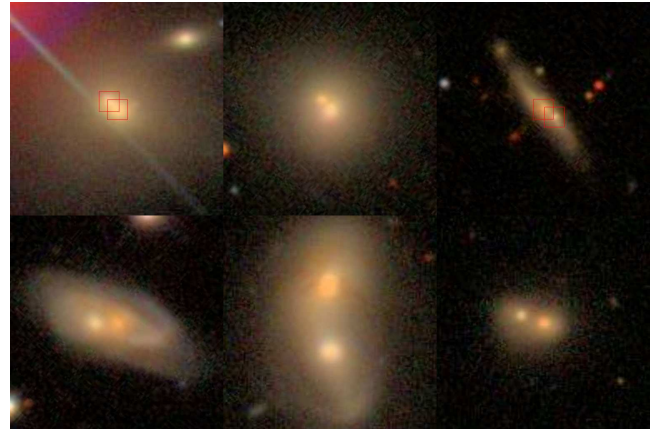}
    \caption{Same as Figure \ref{fig:ambiguous}, but for
    example dumbbell systems in the pair sample. For the systems
    shown, $r_p$ ranges from 2.3 to 9.5 \hseventy\ kpc.
    The long spike in the upper left panel is due to
    diffraction from a nearby star.}
    \label{fig:dumbbell}
\end{figure}

We assign each pair a flag of ``tidal'', ``ambiguous'',
``dumbbell'' (discussed below), or ``non-interaction'' as
listed in Table \ref{table:sample}. We summarize the numbers of
AGNs in each category in Table \ref{table:sum}.  To mitigate
uncertainty of the visual inspection, one of us (X.L.)
classified each object in the pair sample five times, with the
order of objects scrambled each time and not knowing the
results from previous identifications. The identification among
five trials is identical for $\sim80$\% of the objects; for
objects with different classifications from each round (using
``0'' for non-interaction, ``1'' for ambiguous, and ``2'' for
``tidal''; there are few ``dumbbell'' systems and their
classification was almost always the same), we take the median
as the final result. By inspecting the SDSS single-scan images
we found 245 AGN pairs with unambiguous tidal features. In
Figure \ref{fig:best} we show SDSS $gri$-color composite images
of 36 such examples chosen to span the whole range of
separations. The system shown in the third column and second
row of Figure \ref{fig:best} is Mrk 266, which is a luminous
infrared galaxy at $z=0.028$. Its northern nucleus is optically
classified as a Seyfert 2, whereas the southern nucleus is a
LINER. Its double AGN nature is confirmed by {\it Chandra}
X-ray observations (\citealt{mazzarella11}; see also
\citealt{brassington07} and \citealt{wang10}).

\subsubsection{``Ambiguous'' and ``Dumbbell'' Systems}

We excluded from our tidal sample those objects we call
``ambiguous'' pairs of AGNs in which both members exhibit
ordered bar or spiral features but show no clear signs of
galaxy--galaxy interactions.  While these bar or spiral
features could arise from galaxy--galaxy interactions, they may
also be induced by internal instabilities. There are 169
``ambiguous'' pairs and we show six examples in Figure
\ref{fig:ambiguous}. We rejected these pairs because they are
not as convincing as pairs with clear merger-induced tidal
features (such as disruptive asymmetries or tidal bridges).

In addition, we excluded ten ``dumbbell'' systems
\citep[e.g.,][]{valentijn88}, which have double nuclei in a
single host galaxy, but show no optical tidal features
indicative of ongoing interactions between the two nuclei.
Figure \ref{fig:dumbbell} shows six such examples. Three of the
ten systems contain disk components, whereas the others have
optical morphologies reminiscent of 3C75. 3C75 is a double
radio source at the center of the galaxy cluster Abell 400
which shows evidence for interactions only in its twin radio
jets \citep{owen85,lauer88,beers92}.  We did not find evidence
for interactions in the radio by examining the FIRST
\citep{becker95,white97} images of the ``dumbbell'' pairs,
although neither the spatial resolution nor sensitivity is high
enough for a strong constraint.

\begin{figure}
  \centering
    \includegraphics[width=85mm]{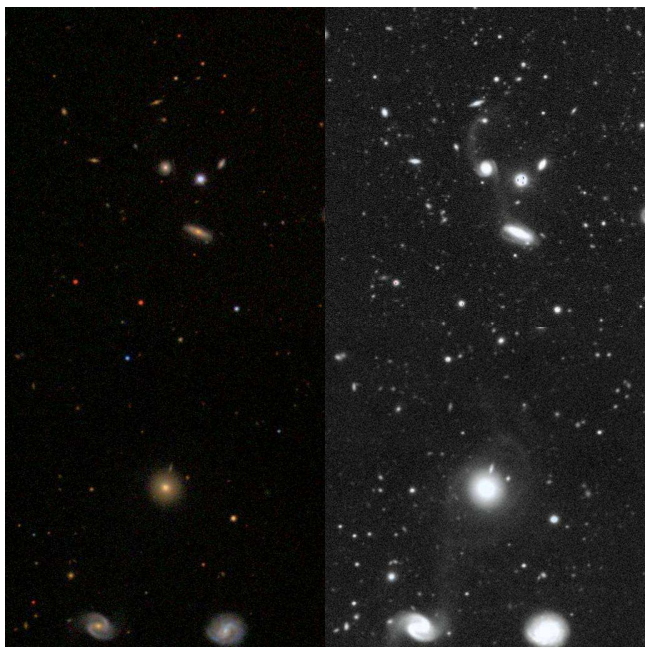}
    \caption{Identifying low surface-brightness tidal features
    using the co-added Stripe 82 images.  We show two examples
    of interacting AGN pairs which show tidal features in the co-added Stripe
    82 images but not in individual SDSS scan images.
    North is up and east is to the left.  The left columns
    are SDSS $gri$-color composite images from individual scans,
    whereas the right columns show the co-added $r$-band
    images in Stripe 82. The FOV of
    each panel is 200\arcsec\ $\times$ 200\arcsec .
    The FOV is centered on one of the AGNs in the pair, which is connected to the
    other galaxy via a tidal stream visible in the co-added images.}
    \label{fig:stripe82}
\end{figure}

\subsubsection{Using the Co-added SDSS Stripe 82 Images to
Identify Low-surface-brightness
Features}\label{subsubsec:stripe82}

Our identification for tidal features is limited by the surface
brightness sensitivity and resolving power of the SDSS images.
The limiting magnitude of individual SDSS scan images with a
signal-to-noise ratio (S/N) of 5 is $r = 23.1$ mag (in the AB
system) for stars \citep{gunn98}.  To quantify the number of
interacting AGNs with LSB tidal features below the sensitivity
of individual SDSS scan images \citep[e.g.,][]{smirnova10}, we
examine the co-added images of a subset of AGNs that are
located in the SDSS Stripe
82\footnote{http://www.sdss.org/legacy/stripe82.html}. Stripe
82 is a sky region covering $\sim270$ deg$^2$ along the
Celestial Equator in the Southern Galactic Cap, which the SDSS
has imaged $\sim100$ times, producing co-added images $\sim$2
mag more sensitive than individual scan images \citep{SDSSDR7}.

We find that $\sim$ 27\% of AGN pairs (6 of 22 pairs) in the
``ambiguous'' category show tidal features in the co-added
Stripe 82 images; the LSB missing fraction is lower ($\sim$
8.5\%, 5 of 58 pairs) for the ``non-interaction'' category.
Figure \ref{fig:stripe82} shows two such examples.  We correct
for the LSB missing fraction using these estimates of the
fraction of the ``ambiguous'' and ``non-interaction'' cases in
what follows.  The LSB correction factor depends on redshift,
since the LSB feature detection is redshift dependent due to
the limited image resolution.  Here what we have estimated
using the co-added images is the effective LSB correction
factor averaged over the redshift range spanned by the sample,
which should be a reasonable approximation as the redshift
distribution of the Stripe 82 sub-sample is similar to the
parent pair sample.  We supplement the tidal AGN pair sample
with the 11 pairs identified using the co-added Stripe 82
images.  Our final AGN pair sample with tidal features consists
of 256 AGN pairs. Table \ref{table:sum} lists the numbers of
AGNs in each category. The fraction of AGN-H {\tiny II}
composites with tidal features (out of all composites in AGN
pairs) is consistent with that of the other AGNs (i.e.,
narrow-line AGNs that satisfy the \citealt{kewley01} criterion
plus those in the other AGN categories as discussed in \S
\ref{sec:selection}) with tidal features (out of all the other
AGNs in pairs). Among the 512 AGNs in the tidal sample, 501
AGNs are contained in the MPA-JHU SDSS DR7 galaxy catalog,
whose diagnostic line ratios are shown in Figure \ref{fig:bpt};
the other 11 objects are selected as either narrow-line or
broad-line quasars.  In the Appendix, we show the SDSS images
and fiber spectra for three examples of AGN pairs with tidal
features.

\begin{figure}
  \centering
    \includegraphics[width=85mm]{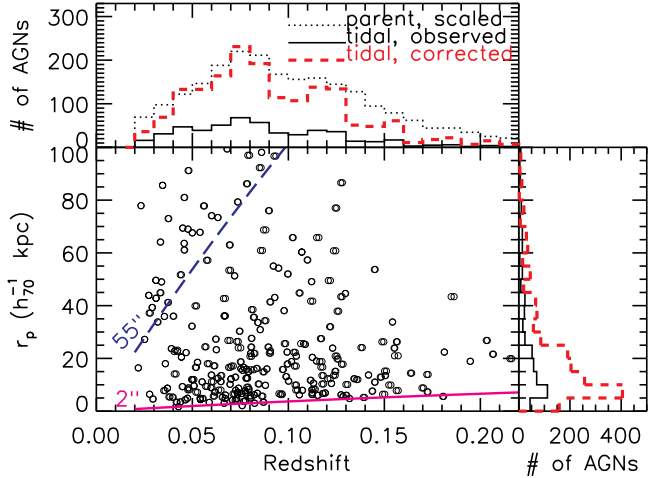}
    \caption{Redshift vs. projected separation ($r_p$) for AGN pairs with tidal features.
    The meanings of the lines are the same as in Figure \ref{fig:rppair}.}
    \label{fig:rpbinary}
\end{figure}

\subsection{Selection Incompleteness}\label{subsec:completeness}

\begin{figure}
  \centering
    \includegraphics[width=85mm]{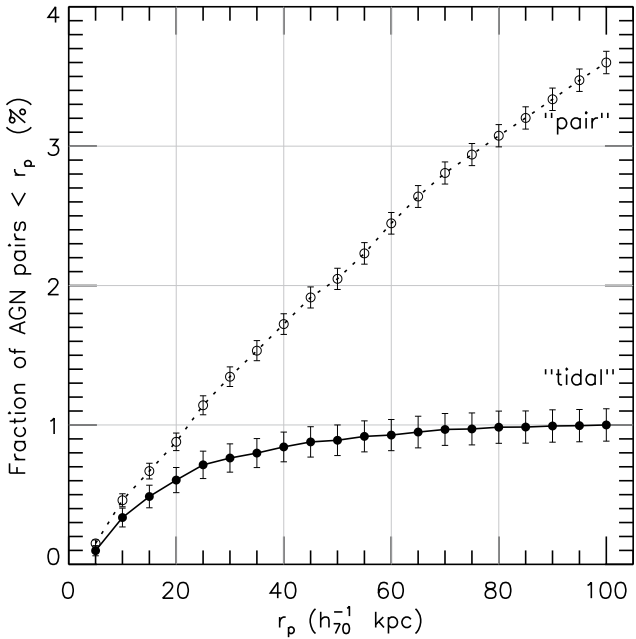}
    \caption{Cumulative fraction of AGN pairs with projected
    separations smaller than $r_p$ (and $\gtrsim 5$ kpc). Open circles denote AGN pairs,
    whereas filled circles represent the subset with tidal features.
    The quoted uncertainties are Poisson errors.}
    \label{fig:rpfraction}
\end{figure}

\begin{figure*}
  \centering
    \includegraphics[width=70mm]{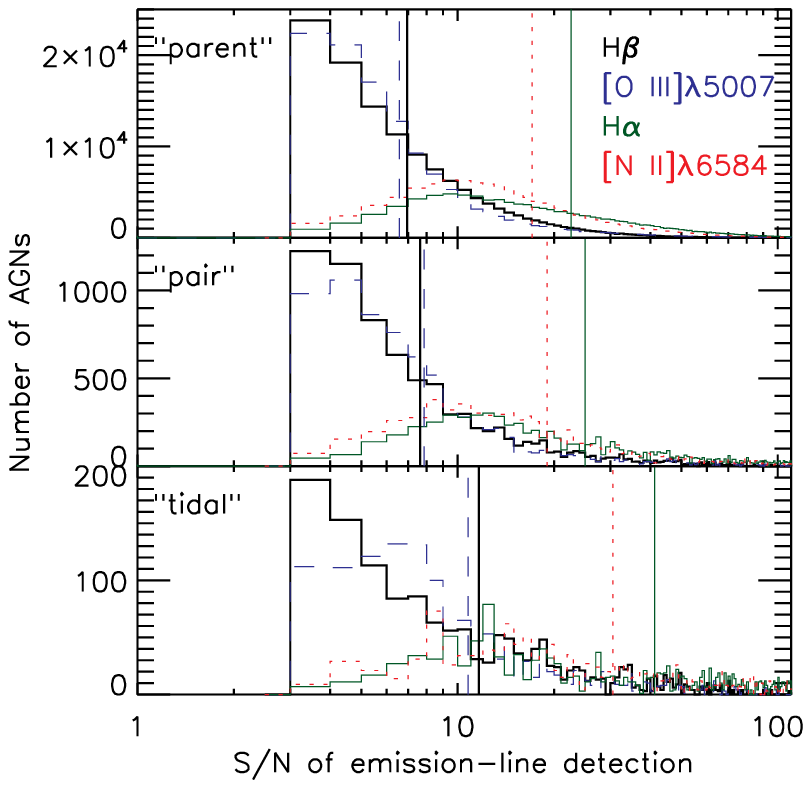}
    \includegraphics[width=70mm]{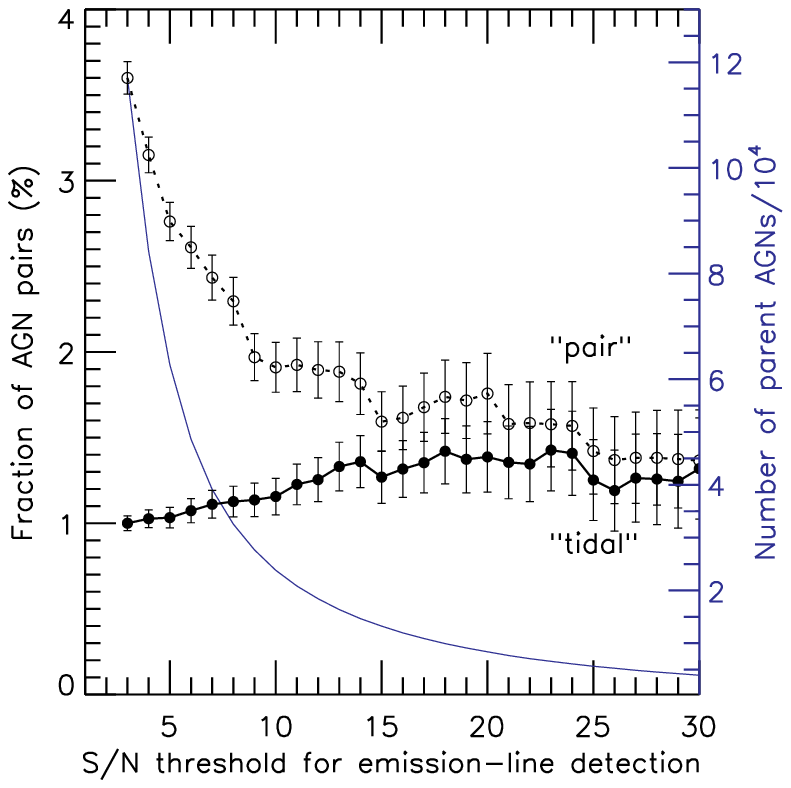}
    \caption{Left: signal-to-noise ratio (S/N) distributions of
    diagnostic emission-line detection of AGNs in
    the parent, pair, and tidal samples (after correction for spectroscopic
    incompleteness).
    Vertical lines indicate the median values for each distribution.
    Right: dependence of the fraction of AGN pairs on the
    S/N threshold adopted for emission-line detection
    (for all four emission lines for both AGNs in a pair).
    Open circles denote AGN pairs, whereas filled circles represent
    the subset with tidal features.
    The quoted uncertainties are Poisson errors.
    The blue solid curve displays the total number
    of AGNs in the parent sample as a function of S/N threshold.}
    \label{fig:s2ndist}
\end{figure*}

The primary factors that limit the sample completeness include
the resolution and surface-brightness limits of SDSS
photometry, fiber collisions, the limitations of visual
identification of tidal features, and the heterogeneity of the
parent sample.

Our approach is insensitive to AGN pairs in advanced mergers
with nuclear separations that are too small to be resolved by
the deblending algorithm of SDSS photometry \citep{lupton01}.
Pairs with nuclear separations $\lesssim 1.''4$ will not be
included. At $z = 0.16$, 1\arcsec\ corresponds to 2.7 kpc in
the assumed cosmology and therefore the sample is incomplete
for $r_p \lesssim 5$ \hseventy\ kpc. At the low-redshift end,
the separation limit is set intrinsically by galaxy size, which
is a function of stellar mass.  AGNs in the pair sample have a
median stellar mass log$(M_{\ast}/M_{\odot})= 10.8$ using the
stellar mass estimates of \citet{kauffmann03c}. The $z$-band
half light radius is observed to be $\sim$3.3 kpc for this
stellar mass \citep{kauffmann03b}.

Our tidal sample does not include mergers at very early stages
before the first pericenter passage; such objects have no
detectable tidal features \citep{toomre72}.  In addition, the
strength of the tidally induced features does not increase
monotonically with decreasing separation when the merger
advances to later stages; it also depends on the progenitor
host-galaxy properties and orbital parameters.  We keep these
separation-dependent limitations and biases in mind when
discussing our results, although we do not attempt to correct
for them. Additional populations of mergers without discernable
optical tidal features may include: minor mergers whose mass
ratios are larger than $\sim$30 \citep{lotz10}, spheroidal
mergers with large bulge-to-disk ratios \citep{mihos96}, and
mergers on significantly retrograde orbits \citep{toomre72}.
Some of these biases are mass or morphology-dependent and
therefore can be corrected for using control samples matched in
galaxy stellar mass (Paper II).

\section{Result: the Frequency of AGN Pairs with $5$ \hseventy\ kpc $\lesssim r_p < 100$
\hseventy\ kpc at $z\sim 0.1$}\label{sec:result}

The fraction of AGN pairs with $5$ \hseventy\ kpc $\lesssim r_p
< 100$ \hseventy\ kpc among the parent AGN sample at $0.02 < z
< 0.16$ is $f_{{\rm p,obs}}\approx$1.1\% (1265 systems out of
117,141 AGNs) before correction for SDSS spectroscopic
incompleteness and is $f_{{\rm p}}\approx$3.6\% after
correction.  As the pair sample may contain false positives due
to closely separated pairs that are not (yet) interacting, this
serves as an upper limit for the intrinsic fraction of
interacting AGN pairs.

Figure \ref{fig:rpbinary} shows redshift versus transverse
proper separation and the distributions of these two quantities
for AGN pairs with tidal features.  There are only nine pairs
at $z> 0.16$.  This apparent deficiency likely results mainly
from the combination of fewer objects in the parent sample and
the decreasing apparent size of tidal features with redshift.
The redshift distribution of the tidal sample is roughly
consistent with the parent sample both before and after
correction for fiber collisions, except that the tidal sample
is incomplete for smaller separation pairs (projected angular
separation $\lesssim 2''$).  The number of tidal pairs found
increases with decreasing $r_p$ (except for the bin at the
smallest scales, which is affected by the resolution limit of
SDSS photometry), a trend that is partly caused by the fact
that closer pairs tend to show more prominent tidal features.

We estimate the fraction of AGN pairs with tidal features among
the parent AGNs at $0.02 < z < 0.16$ as
\begin{equation}
f_{{\rm t}} = C(f_{{\rm lap}}, p)f_{{\rm t,obs}} = C(f_{{\rm lap}}, p)\frac{(N_{t} + C_a
N_a + C_n N_n)}{2N},
\end{equation}
where $C(f_{{\rm lap}}, p) \approx 3.3$ is the correction
factor for spectroscopic incompleteness (Equation
\ref{eq:fiber}), $N_t$ is the number of galaxies in AGN pairs
with tidal features identified in individual SDSS images, $N_a$
and $N_n$ are the numbers of galaxies classified as
``ambiguous'' and ``non-interaction'', $C_a = 27$\% and $C_n =
8.5$\% are the correction factors for LSB features estimated
based on our Stripe 82 experiment (\S
\ref{subsubsec:stripe82}), and $N$ is the number of AGNs in the
parent sample.  We find $f_{{\rm t,obs}}\approx$0.3\% before
correction for spectroscopic incompleteness and $f_{{\rm
t}}\approx$1.0\% after correction. Because the tidal sample is
incomplete for interacting AGN pairs without detectable tidal
features, this serves as a lower limit for the intrinsic
fraction of interacting AGN pairs.

Figure \ref{fig:rpfraction} presents how our results vary as a
function of upper limit of $r_p$. Observations of statistical
samples of inactive galaxy pairs suggest that $\sim30$
\hseventy\ kpc is the physical scale in projected separation
below which galaxy pairs exhibit significantly higher star
formation rates than field galaxies
\citep[e.g.,][]{barton00,lambas03,alonso04,nikolic04}. The
fraction of AGN pairs with $5$ \hseventy\ kpc $\lesssim r_p <
30$ \hseventy\ kpc is 1.3\%, among which $\sim60$\% show tidal
features.

The fraction of AGN pairs we measure depends on the sensitivity
of our AGN identification. We adopted an S/N of $3$ as our
baseline value for the detection threshold of the diagnostic
emission lines, \hbeta , \OIIIb , \halpha , and \NIIb, in the
selection of narrow-line AGNs. We now address how the S/N
threshold affects our results for the fraction of AGN pairs.
The sample statistics are dominated by narrow-line AGNs (Table
\ref{table:sum}), and we thus neglect the effects of changing
the selection criteria for the other AGN categories. In the
left panel of Figure \ref{fig:s2ndist}, we show the S/N
distribution of the four diagnostic emission lines, for AGNs in
the parent, pair, and tidal samples (both corrected for fiber
incompleteness for the latter two), respectively. For these
four emission lines, AGNs in the pair sample have similar S/N
distributions to those of the parent AGN sample. On the other
hand, AGNs in the tidal sample have larger emission-line S/N
than do the parent AGNs; tidal AGNs have median emission-line
S/N which are $\sim$1.5 times those of parent AGNs. This may be
explained if AGN pairs with more prominent tidal features tend
to exhibit stronger emission lines (and thereby have higher S/N
detections).  We increase the S/N threshold and redo the
analysis.  In the right panel of Figure \ref{fig:s2ndist}, we
show how the fraction of AGN pairs (and those with tidal
features) varies with increasing S/N threshold.   The fraction
of AGN pairs drops from $\sim3.6$\% at S/N$>3$ to $\sim1.4$\%
at S/N$\gtrsim30$. Because the pair and parent samples have
similar S/N distributions, the pair fraction would scale
$\propto N^2/N\propto N$ if AGNs were uncorrelated (i.e.,
randomly paired), where $N$ denotes the total number of parent
AGNs. The observed decay of pair fraction with decreasing $N$
is less steep than the uncorrelated case would predict. The
fraction of AGN pairs with detected tidal features roughly
stays constant with increasing S/N threshold, although we
caution that it should be treated as a lower limit and the
incompleteness due to the inspection of tidal features is
likely higher at lower S/N, if AGN pairs with more prominent
tidal features tend to exhibit stronger emission lines. At
S/N$\gtrsim 15$, $\gtrsim80$\% of all AGN pairs show tidal
features (Figure \ref{fig:s2ndist}).

\section{Discussion}\label{sec:discuss}

\subsection{Implications for AGN Pairs with Smaller Separations}\label{subsec:fraction}

The fraction of AGN pairs at $0.02< z < 0.16$ with $5$
\hseventy\ kpc $\lesssim r_p < 100$ \hseventy\ kpc is
$\sim$3.6\% corrected for SDSS spectroscopic incompleteness.
The pair fraction decreases to $\sim$1.0\% if we restrict
ourselves to those pairs that show clear tidal features in SDSS
images.  If we assume that (1) the fraction of observed pairs
within a certain range of projected separations scales linearly
with the time $\tau$ that a merger spends in that range, and
(2) the probability that the two AGNs in a merger are
simultaneously active does not strongly depend on the merging
phase over the range of separations we are considering (i.e., a
few kpc to tens-of-kpc scales), then the fraction of AGN pairs
on tens-of-kpc scales can be related to the fraction on kpc
scales as $f_{{\rm kpc}} \approx f_{{\rm
tens-of-kpc}}\tau_{{\rm kpc}}/\tau_{{\rm tens-of-kpc}}$. The
timescales $\tau$ are proportional to the length scales, so
$f_{{\rm kpc}}$ will be $\sim$0.1\% or $\sim$0.4\% using the
tidal and full pair samples, respectively.

Several studies \citep{liu10,smith09,wang09} have found that
$\sim 1$\% of SDSS AGNs show double-peaked profiles in \OIIIc\
emission lines, in which the two velocity peaks are blueshifted
and redshifted from the systemic velocity determined from
stellar absorption lines.  As shown in \citet{shen10b}, at
least $\sim$50\% of the double-peaked narrow emission lines are
due to narrow-line region kinematics such as biconical outflows
or rotating disks
\citep[e.g.,][]{axon98,veilleux01,crenshaw09}, while $\sim$10\%
or more reflect the orbital motion of a merging pair of AGNs
\citep[e.g.][]{zhou04,comerford08}. In the merging AGN pair
scenario for those double-peaked AGNs, the projected angular
separation of the two nuclei has to be smaller than 3$''$ for
both AGNs to be covered by a single SDSS fiber. This
corresponds to $r_p < 8$ \hseventy\ kpc for the typical
redshift of $z \sim 0.15$ in the \citet{liu10} sample. The
fraction of AGNs in kpc-scale pairs we have inferred ($f_{{\rm
kpc}}\sim$ 0.1\%--0.4\%) suggests that only $\sim$ 10\%--40\%
of AGNs with double-peaked narrow emission lines are merging
systems. However, this should be treated as an upper limit,
considering there are kpc-scale AGN pairs with LOS velocity
offsets $<150$ km s$^{-1}$ and therefore will not be identified
as SDSS double-peaked narrow-line AGNs.  We caution that the
estimate is highly uncertain since $\tau_{{\rm kpc}}/\tau_{{\rm
tens-of-kpc}}$ depends on merger parameters and the host-galaxy
properties of AGN pairs. In addition, we use the
\citet{kauffmann03} criterion for diagnostic line ratios,
whereas the \citet{kewley01} criterion was adopted in the
\citet{liu10} sample, whose S/N cut for spectral measurements
was also more stringent. Nevertheless, \citet{liu10b} used
follow-up observations to confirm four kpc-scale binary AGNs
out of 43 double-peaked objects of the \citet{liu10} sample
\citep[see also][]{shen10b}, in broad agreement with our
estimate of $f_{{\rm kpc}}$ based on AGN pairs with broader
separations.

\subsection{Comparison with Binary Quasars}\label{subsec:bqso}

The fraction of quasars in pairs (with bolometric luminosities
of $\gtrsim 10^{46}$ erg s$^{-1}$) with separations of tens to
hundreds of kpc is $\lesssim 0.1$\% at $1 < z <5$
\citep[e.g.,][]{hennawi06,hennawi09,myers08,shen10c}. This is
$\sim$8--30 times smaller than what we observe for the less
luminous AGN pairs at $\bar{z}\sim 0.08$.  If we assume that
(1) all luminous high-redshift quasars are triggered in
galaxy-galaxy mergers, and (2) the two quasars in a pair shine
at random, uncorrelated times, among all quasars the observed
fraction of quasar pairs with separations of tens to hundreds
of kpc will be roughly the ratio of the quasar lifetime $
\tau_{{\rm quasar}}$ and the time for two galaxies with this
separation to merge $\tau_{{\rm merge}}$. Both $\tau_{{\rm
quasar}}$ and $\tau_{{\rm merge}}$ span orders of magnitude in
their parameter spaces. We speculate that the first of the two
assumptions is likely valid for high-redshift luminous quasars,
as mergers are arguably the most efficient mechanism to give
rise to such high-mass accretion rates.  The second assumption
is valid, provided that the separations of observed quasar
pairs (only a few of which are tens-of-kpc separations) are too
large for galaxy--galaxy tidal interactions to be effective
(the counterexample of an interacting quasar pair reported by
\citealt{green10} has a separation of 21 kpc which is at the
lower bound of known quasar pairs). However, these two
assumptions do not necessarily apply to the less luminous AGN
pairs; their intrinsic accretion luminosities are roughly three
orders of magnitude lower, and the typical projected
separations in our sample are smaller by a factor of $\sim$10
than the high-redshift luminous quasar pairs. Our AGN pair
sample (low-luminosity, low-redshift) and the (luminous,
high-redshift) binary quasar samples probe populations at very
different redshift and AGN-luminosity regimes, such that a
useful direct comparison with the binary quasar samples cannot
be made within our sample. In particular, the statistics of our
sample is dominated by Seyfert galaxies (and mostly narrow-line
Seyferts; Table \ref{table:sum}).

\subsection{What Fraction of Moderate-luminosity AGNs Are Triggered in Galaxy
Interactions?}\label{subsec:merger}

As above, if we assume that all moderate-luminosity AGNs were
triggered in galaxy interactions and the two components in a
pair shone at uncorrelated times, the fraction of AGNs in pairs
with $5$ \hseventy\ kpc $ < r_p < 100$ \hseventy\ kpc would
roughly be the ratio between the typical lifetime of
moderate-luminosity AGNs $\tau_{{\rm AGN}}$ and the dynamical
timescale of galaxy mergers with $5$ \hseventy\ kpc $ < r_p <
100$ \hseventy\ kpc $\tau_{5-100\,\, {\rm kpc}}$, which is
$\sim$ 10\% assuming $\tau_{{\rm AGN}}\sim 10^8$ yr and
$\tau_{5-100\,\, {\rm kpc}}\sim10^9$ yr, larger than the $\sim$
1\%--4\% value that we have found.  This implies that only
$\sim 10$\%--$40$\%$\times(10^8 \,{\rm yr}/\tau_{{\rm
AGN}})(\tau_{5-100\,\, {\rm kpc}}/10^9 \,{\rm yr})$ of
moderate-luminosity AGNs can be triggered in galaxy
interactions.  In addition, we will show in Paper II that the
strengths of AGN activity are correlated between the two
components in interacting pairs.  This correlation suggests
that the two SMBHs in a galaxy pair may tend to be activated
simultaneously, making the expected fraction of AGN pairs even
larger than $\sim \tau_{{\rm AGN}}/\tau_{5-100\,\, {\rm kpc}}$.
Thus, the observed fraction of AGN pairs suggests that fewer
than $\sim 10$\%--$40$\%$\times(10^8 \,{\rm yr}/\tau_{{\rm
AGN}})(\tau_{5-100\,\, {\rm kpc}}/10^9 \,{\rm yr})$ of
moderate-luminosity AGNs are triggered in galaxy interactions.

\section{Summary}\label{sec:sum}

We have selected a sample of 1286 AGN pairs (or multiples) with
LOS velocity offsets $< 600$ km s$^{-1}$ and projected
separations $< 100$ \hseventy\ kpc from 138,070 optical AGNs in
SDSS DR7. Two hundred fifty-six pairs of these show unambiguous
morphological tidal features in their SDSS images indicative of
ongoing interactions.  After correction for spectroscopic
incompleteness, the fraction of AGN pairs with $5$ \hseventy\
kpc $\lesssim r_p \leq 100$ \hseventy\ kpc and $\Delta v < 600$
km s$^{-1}$ is $\sim$3.6\% among the parent AGNs at $0.02 < z
<0.16$. The fraction of AGN pairs that show tidal features is
$\sim$1.0\%.

The current study is a part of our continuing effort to
systematically identify and characterize the populations of
merging SMBHs at various stages. The sample that we have
presented increases the number of known AGN pairs on $\sim
5$--$100$ \hseventy\ kpc scales by more than an order of
magnitude, thanks to the statistical power of SDSS. It
constitutes the starting point of a statistical analysis of
their properties. In Paper II, we will examine the effects of
tidal interactions on AGN pairs by quantifying their recent
star formation and BH accretion activity as a function of pair
separation, calibrated against a control sample of AGNs matched
in both redshift and stellar mass distribution, and we will
examine correlations between the interacting components.

\acknowledgments

We thank J. Greene and P. Kampczyk for helpful comments, and an
anonymous referee for a prompt and careful report.  We are
grateful to the MPA-JHU team, especially J. Brinchmann, who
have made their catalog of derived galaxy properties publicly
available. X.L. and M.A.S. acknowledge the support of NSF grant
AST-0707266. Support for the work of X.L. was provided by NASA
through Einstein Postdoctoral Fellowship grant number
PF0-110076 awarded by the Chandra X-ray Center, which is
operated by the Smithsonian Astrophysical Observatory for NASA
under contract NAS8-03060. Y.S. acknowledges support from a
Clay Postdoctoral Fellowship through the Smithsonian
Astrophysical Observatory.

Funding for the SDSS and SDSS-II has been provided by the
Alfred P. Sloan Foundation, the Participating Institutions, the
National Science Foundation, the U.S. Department of Energy, the
National Aeronautics and Space Administration, the Japanese
Monbukagakusho, the Max Planck Society, and the Higher
Education Funding Council for England. The SDSS Web site is
http://www.sdss.org/.

The SDSS is managed by the Astrophysical Research Consortium for
the Participating Institutions. The Participating Institutions are
the American Museum of Natural History, Astrophysical Institute
Potsdam, University of Basel, University of Cambridge, Case
Western Reserve University, University of Chicago, Drexel
University, Fermilab, the Institute for Advanced Study, the Japan
Participation Group, Johns Hopkins University, the Joint Institute
for Nuclear Astrophysics, the Kavli Institute for Particle
Astrophysics and Cosmology, the Korean Scientist Group, the
Chinese Academy of Sciences (LAMOST), Los Alamos National
Laboratory, the Max-Planck-Institute for Astronomy (MPIA), the
Max-Planck-Institute for Astrophysics (MPA), New Mexico State
University, Ohio State University, University of Pittsburgh,
University of Portsmouth, Princeton University, the United States
Naval Observatory, and the University of Washington.

{\it Facility}: Sloan



\appendix

\section{Examples of AGN Pairs with Tidal Features}

\begin{figure}
  \centering
    \includegraphics[width=53mm]{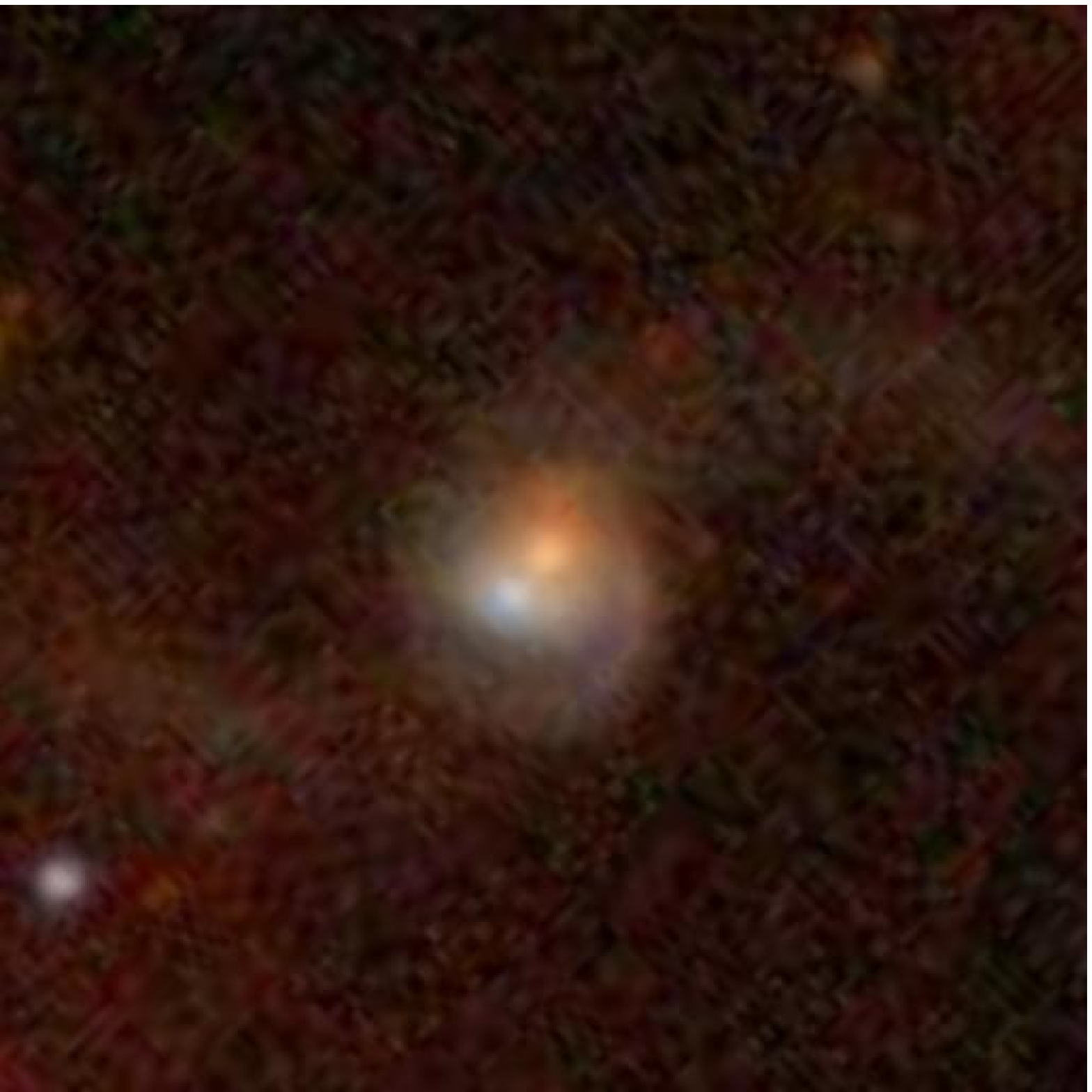}
    \includegraphics[width=110mm]{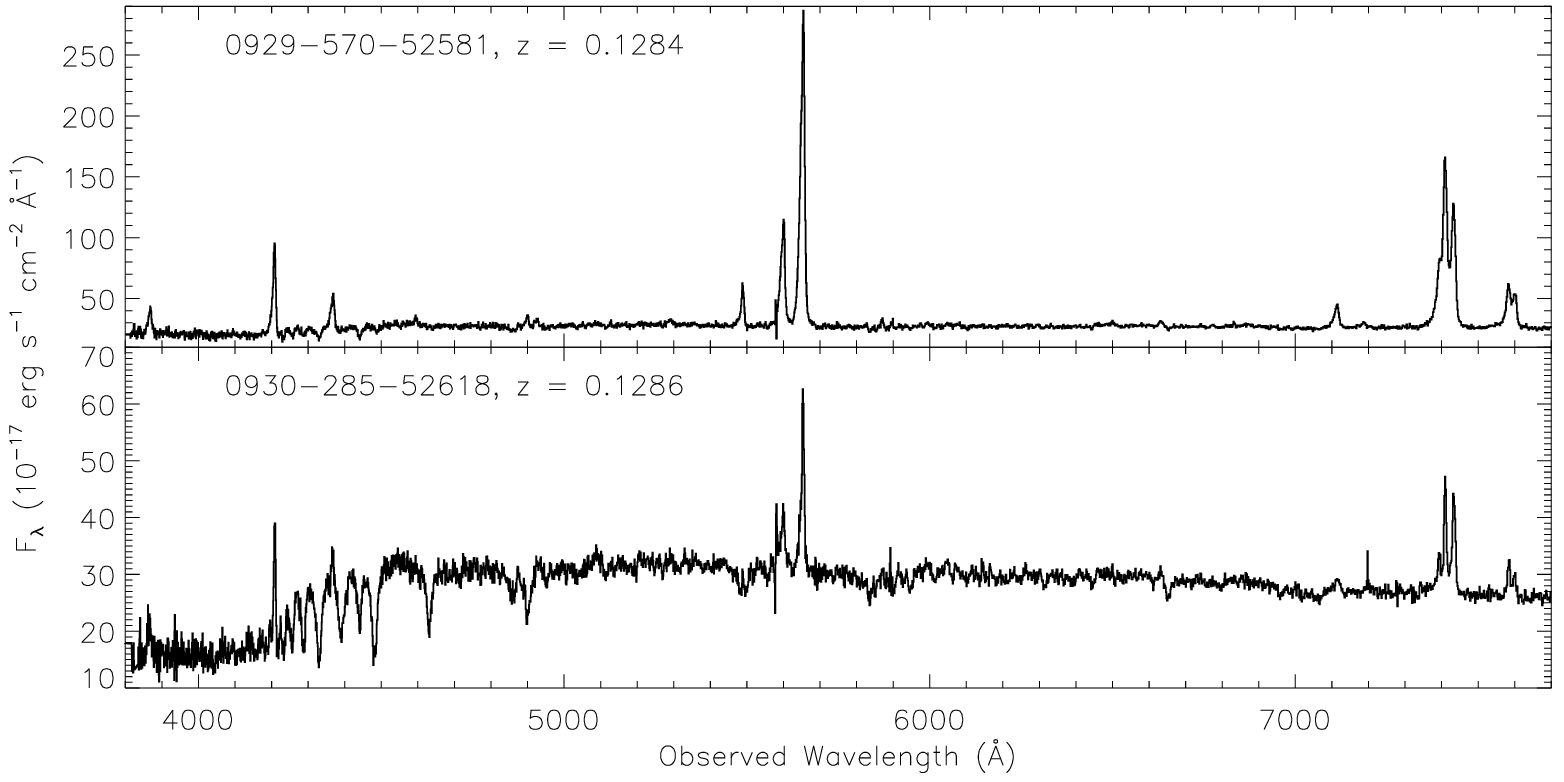}
    \includegraphics[width=53mm]{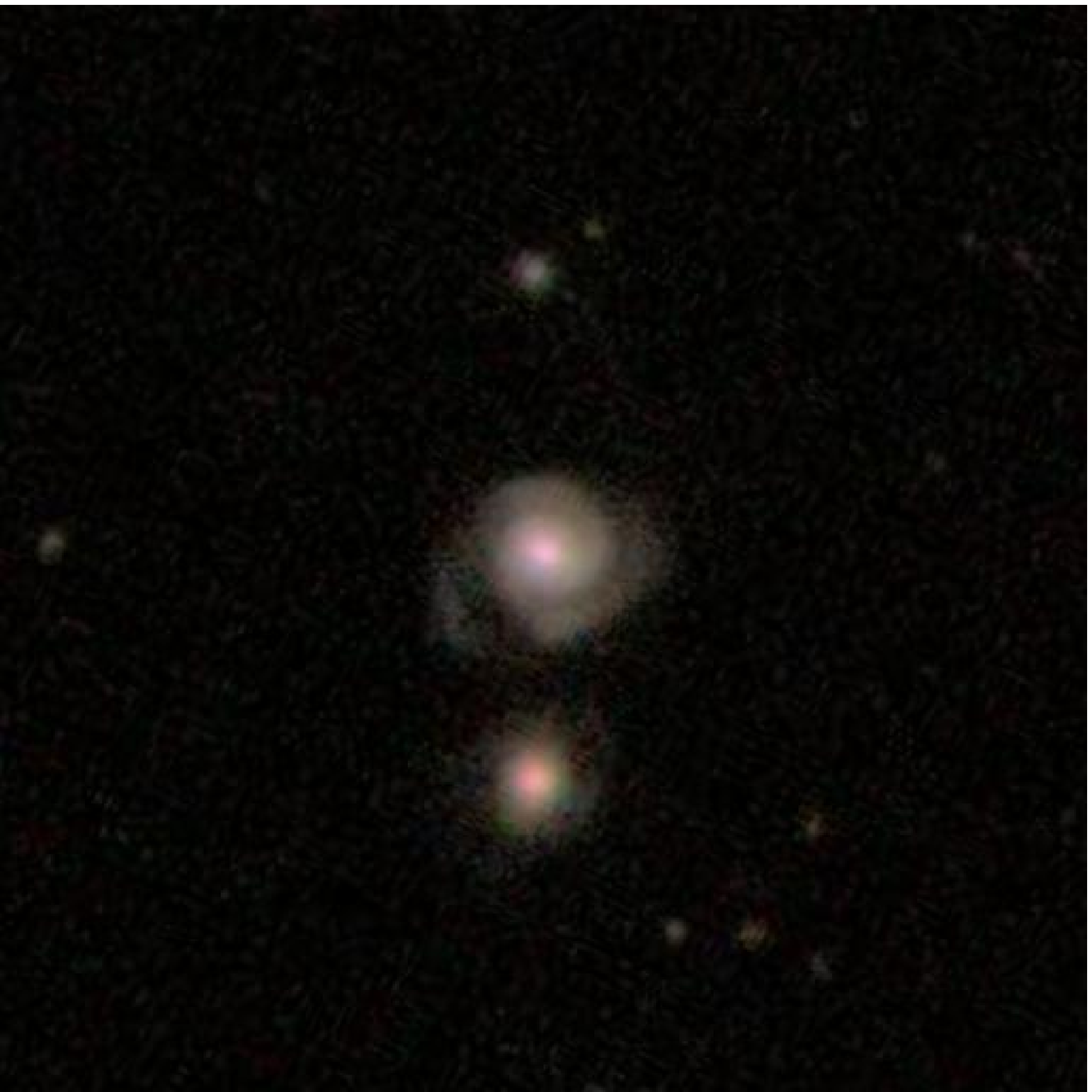}
    \includegraphics[width=110mm]{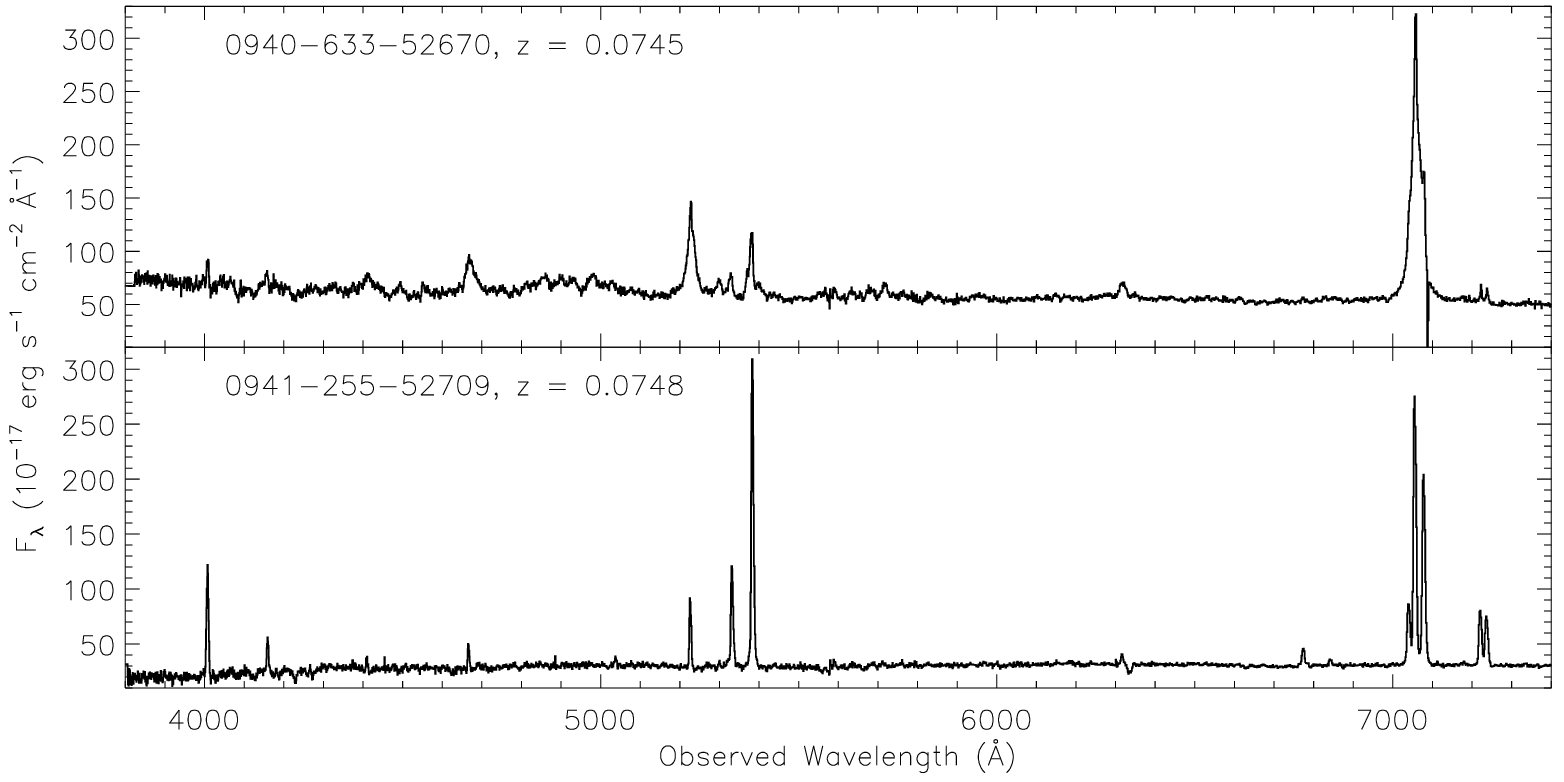}
    \includegraphics[width=53mm]{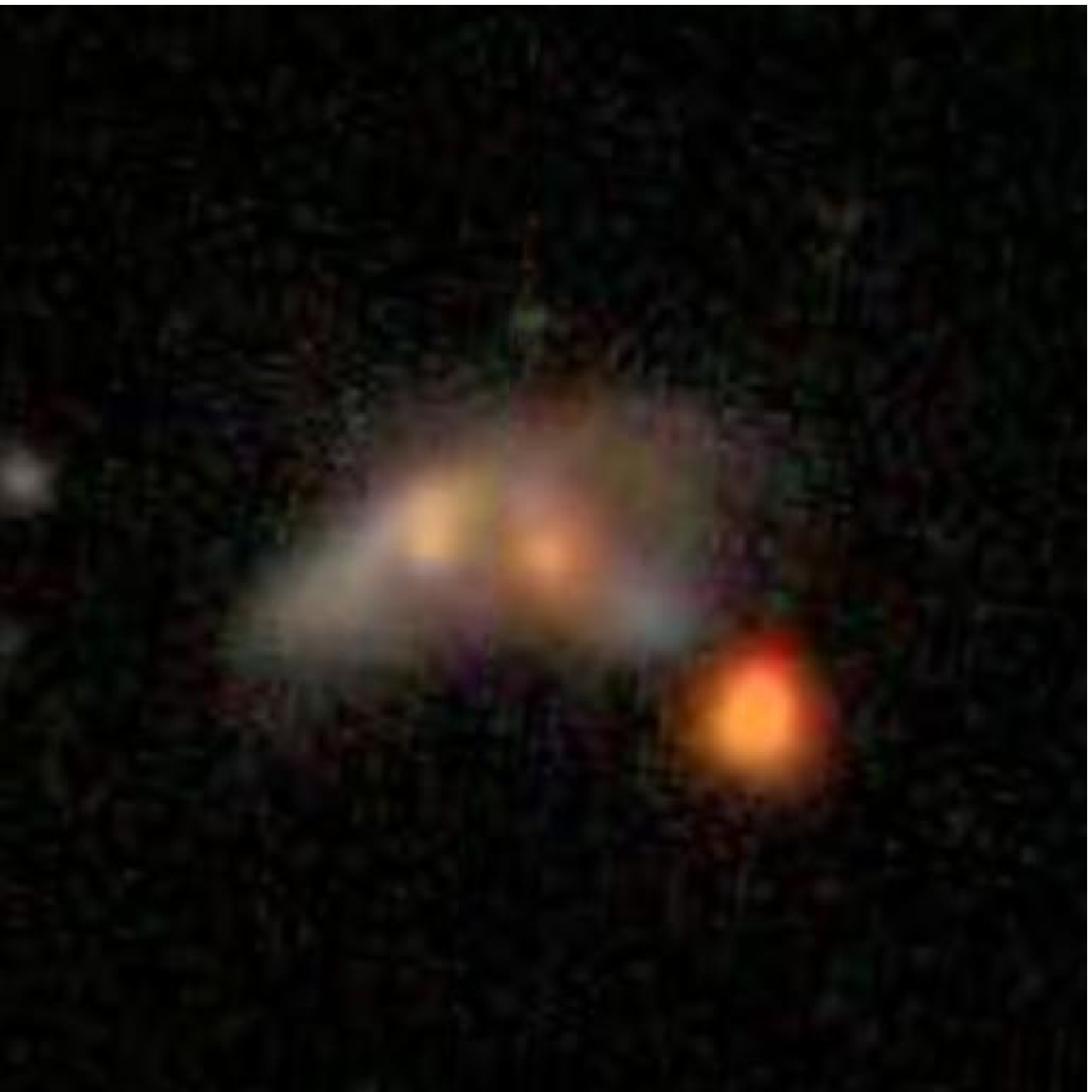}
    \includegraphics[width=110mm]{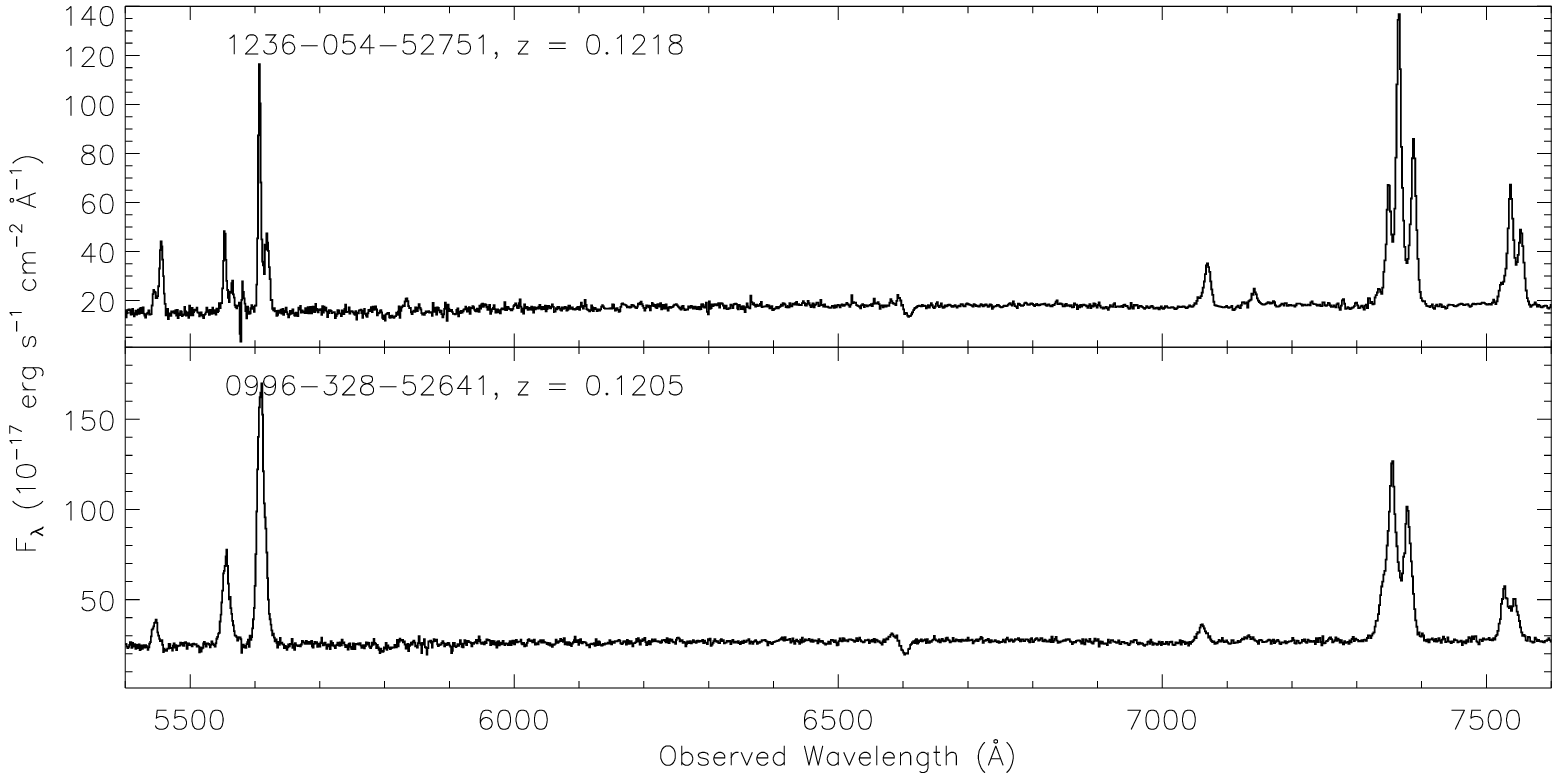}
    \caption{Three examples of AGN pairs with tidal features.
     Left column: SDSS $gri$-color composite images.
     North is up and east is to the left.
     FOV is $50''\times50''$ for the top and bottom objects, and
     $100''\times100''$ for the middle one.
     Right column: SDSS fiber spectra.
     Labeled on each panel are
     spectroscopic plate number, fiber ID, and modified Julian date.
     Top: 0929-570-52581 is the NW component,
     and 0930-285-52618 is the SE component.
     The two AGNs are separated by $2.''4$ (5.6 \hseventy\ kpc).
     Middle:  the northern component (940-633-52670) is a broad-line,
     and the southern component (941-255-52709) is a narrow-line AGN.
     They are separated by $21.''3$ (30.3 \hseventy\ kpc).
     Bottom: 1236-54-52751 is the western component which
     has double-peaked narrow-emission lines, and 996-328-52641 is
     the eastern component.
     The two AGNs are separated by $5.''7$ (12.3 \hseventy\ kpc).
     The object to the southwest is a foreground star.}
    \label{fig:egspec}
\end{figure}

In Figure \ref{fig:egspec}, we show SDSS images and spectra for
three examples of AGN pairs with tidal features.  Note that for
AGN pairs with small (kpc-scale) separations (e.g., the top
object shown in Figure \ref{fig:egspec}), we cannot rule out
the possibility that one AGN is ionizing both galaxies
\citep[e.g.,][]{liu10b}. High-resolution X-ray and/or radio
observations can help better constrain the double or single AGN
nature of these sources.

\bibliography{binaryrefs}

\end{document}